\documentclass{aa}
\usepackage{graphicx}
\usepackage{natbib}

\def\hh{H$_2$}
\def\hhh{H$_3^+$}
\def\hhdp{H$_2$D$^+$}
\def\nnhp{N$_2$H$^+$}
\def\nndp{N$_2$D$^+$}
\def\ammo{NH$_3$}
\def\dammo{ND$_3$}
\def\nhhd{NH$_2$D}
\def\nddh{ND$_2$H}

\def\ltsim{{_<\atop{^\sim}}}
\def\kms{km~s$^{-1}$}
\def\rcm{cm$^{-1}$}
\def\scm{cm$^{-2}$}
\def\ccm{cm$^{-3}$}
\def\tkin{$T_{\rm kin}$}
\def\tex{$T_{\rm ex}$}
\def\tmb{$T_{\rm mb}$}
\def\tsys{$T_{\rm sys}$}

\begin{document}
\title{Triply deuterated ammonia in NGC 1333} 

\author{F.F.S.\ van der Tak\inst{1} \and P.\ Schilke\inst{1} \and
  H.S.P.\ M\"uller\inst{2} \and D.C.\ Lis\inst{3} \and T.G.\ 
  Phillips\inst{3}\and M.\ Gerin\inst{4,5}\and E.\ Roueff\inst{5}} 

\offprints{vdtak@mpifr-bonn.mpg.de} 

\institute{Max-Planck-Institut f\"ur Radioastronomie, Auf dem H\"ugel
  69, 53121 Bonn, Germany \and I.~Physikalisches Institut,
  Universit\"at zu K\"oln, 50937 K\"oln, Germany \and California
  Institute of Technology, Downs Laboratory of Physics 320-47,
  Pasadena, CA 91125, USA \and Lab.\ de Radioastronomie
  Millim\'etrique, D\'ept.\ de Physique de l'E.N.S., 24 Rue
  Lhomond, 75231 Paris, France \and DEMIRM, Observatoire de Paris, 61
  avenue de l'Observatoire, 75014 Paris, France}

\date{Received April 1, 2002; accepted April 24, 2002}

\abstract{The Caltech Submillimeter Observatory has detected triply
  deuterated ammonia, \dammo, through its $J_K=1_0^a\to0_0^s$
  transition near 310~GHz. Emission is found in the NGC~1333 region,
  both towards IRAS 4A and a position to the South-East where DCO$^+$
  peaks. In both cases, the hyperfine ratio indicates that the
  emission is optically thin. Column densities of \dammo\ are
  $3-6\times 10^{11}$~\scm\ for \tex=10~K and twice as high for
  \tex=5~K.  Using a Monte Carlo radiative transfer code and a model
  of the structure of the IRAS source with temperature and density
  gradients, the estimated \dammo\ abundance is $3.2\times 10^{-12}$
  if \dammo/\hh\ is constant throughout the envelope. In the more
  likely case that \dammo/\hhdp\ is constant, \dammo/\hh\ peaks in the
  cold outer parts of the source at a value of $1.0\times
  10^{-11}$. To reproduce the observed \ammo/\dammo\ abundance ratio
  of $\sim$1000, grain surface chemistry requires an atomic D/H ratio
  of $\approx$0.15 in the gas phase, $>$10 times higher than in recent
  chemical models. More likely, the deuteration of \ammo\ occurs by
  ion-molecule reactions in the gas phase, in which case the data
  indicate that deuteron transfer reactions are much faster than
  proton transfers.}

\maketitle


\section{Introduction}
\label{sec:intro}

Deuterium-bearing molecules have attracted attention in recent years.
Physically, they appear to be good probes of the very cold phases of
molecular clouds prior to star formation. Chemically, the isotopic
composition of molecules is an important clue to their formation
mechanism. There are two ways to make deuterated molecules. First, at
temperatures $\ltsim 70$~K, the gas-phase reaction equilibrium ${\rm
  H}_3^+ + {\rm HD} \rightleftharpoons {\rm H}_2{\rm D}^+ + {\rm H}_2$
is shifted in the forward direction. Subsequent deuteron transfer from
H$_2$D$^+$ to, e.g., CO and N$_2$, leads to the large observed
abundance ratios of DCO$^+$/HCO$^+$ and N$_2$D$^+$/\nnhp\ of
$\sim$0.1, four orders of magnitude higher than the Galactic D/H ratio
\citep{turn01}.  The key species of this chemical scheme, \hhdp, was
recently detected in the Class~0 source NGC~1333 IRAS~4A
\citep{stark99}. The high densities and low temperatures in this
object promote the formation of \hhdp\ out of \hhh\ and HD, and also
prevent its destruction because the major destroyer of \hhdp, CO, is
depleted by a factor of $\sim$100 due to freeze-out onto dust grains.

Alternatively, deuterium-bearing molecules can be formed on dust
grains by surface chemistry. Accretion of H and CO onto grains,
followed by reaction, is thought to produce solid H$_2$CO and
CH$_3$OH. This mechanism favours deuteration because the atomic D/H
ratio in the gas phase is much greater than the elemental ratio.  The
observed abundances of HDCO, CH$_3$OD and D$_2$CO in Orion and IRAS
16293 indicate their synthesis on dust grains
\citep{turn90,char97,cecc98}. The same mechanism may work for \ammo,
provided most nitrogen is in atomic form, and support for this route
comes from detections of solid \ammo\ \citep{lacy98,gibb00}.  However,
these detections remain tentative \citep{dart01}, and observations of
\nnhp\ suggest that in dense clouds, most nitrogen is in molecular
form \citep{woma92}.

While \nhhd\ has been observed in many sources \citep{sait00,shah01},
\nddh\ has only been detected so far towards the cold, starless cores
L134N and L1689N \citep{roueff00,loin01}.  The temperatures in these
sources of $\approx$10~K are too low for significant evaporation of
even the most volatile ices to occur.  Gas-phase reactions can
probably account for the observed abundance ratios \citep{rodg01}, but
observations of \dammo\ would present a strong test, as the gas-phase
route produces $\approx$3 times more \dammo\ than the grain surface
route. As a step towards measuring the relative importance of
gas-phase and solid-state deuteration, we have observed the
$J_K=1_0\to0_0$ transition of \dammo\ towards NGC~1333.
Together with similar observations towards Barnard~1 by \citet{lis02},
this is the first detection of a triply deuterated molecule in
interstellar space.

\section{Observations and Results}
\label{sec:obs}

The rotational energy levels of symmetric top molecules are labeled by
the total angular momentum $J$ and its projection on the molecular
symmetry axis $K$. For \ammo\ and \dammo, inversion motion splits each
level further into states which are symmetric (s) and antisymmetric
(a) upon reflection in the plane of the H~or D~atoms.  The measured
frequencies of the \dammo\ $J_K=1_0^a\to0_0^s$ transition are
309908.46 ($F$=1$\to$1), 309909.69 ($F$=2$\to$1) and 309911.53
($F$=0$\to$1) MHz; those of the $J_K=1_0^s\to0_0^a$ transition are
306735.58 ($F$=1$\to$1), 306736.96 ($F$=2$\to$1) and 306738.95
($F$=0$\to$1) MHz \citep{helm69}. Due to spin statistics, the
$J_K=1_0^a\to0_0^s$ transition is stronger than the
$J_K=1_0^s\to0_0^a$ transition by a factor of 10.

\begin{figure}[t]
\resizebox{\hsize}{!}{\includegraphics[angle=-90]{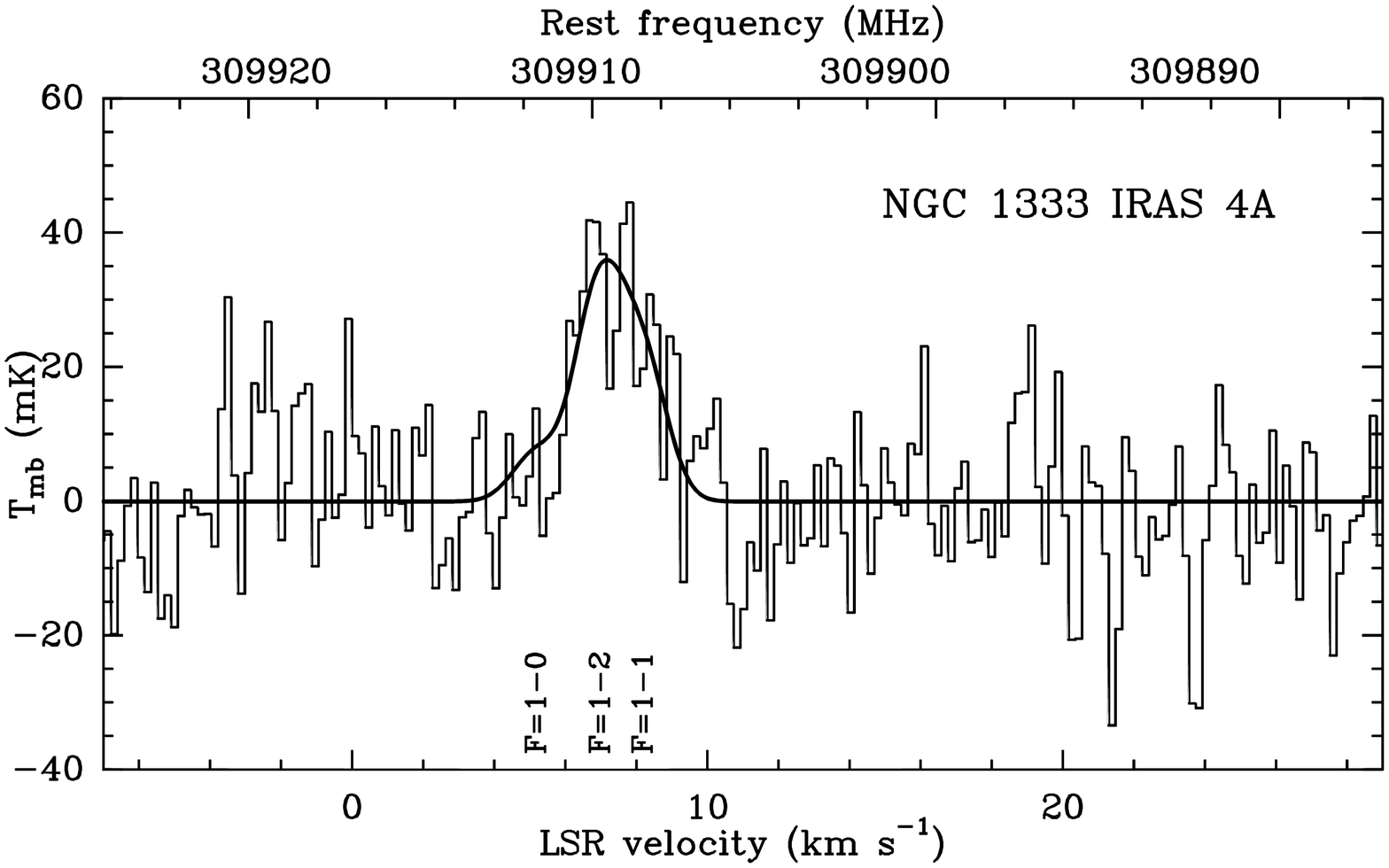}}
\bigskip
\resizebox{\hsize}{!}{\includegraphics[angle=-90]{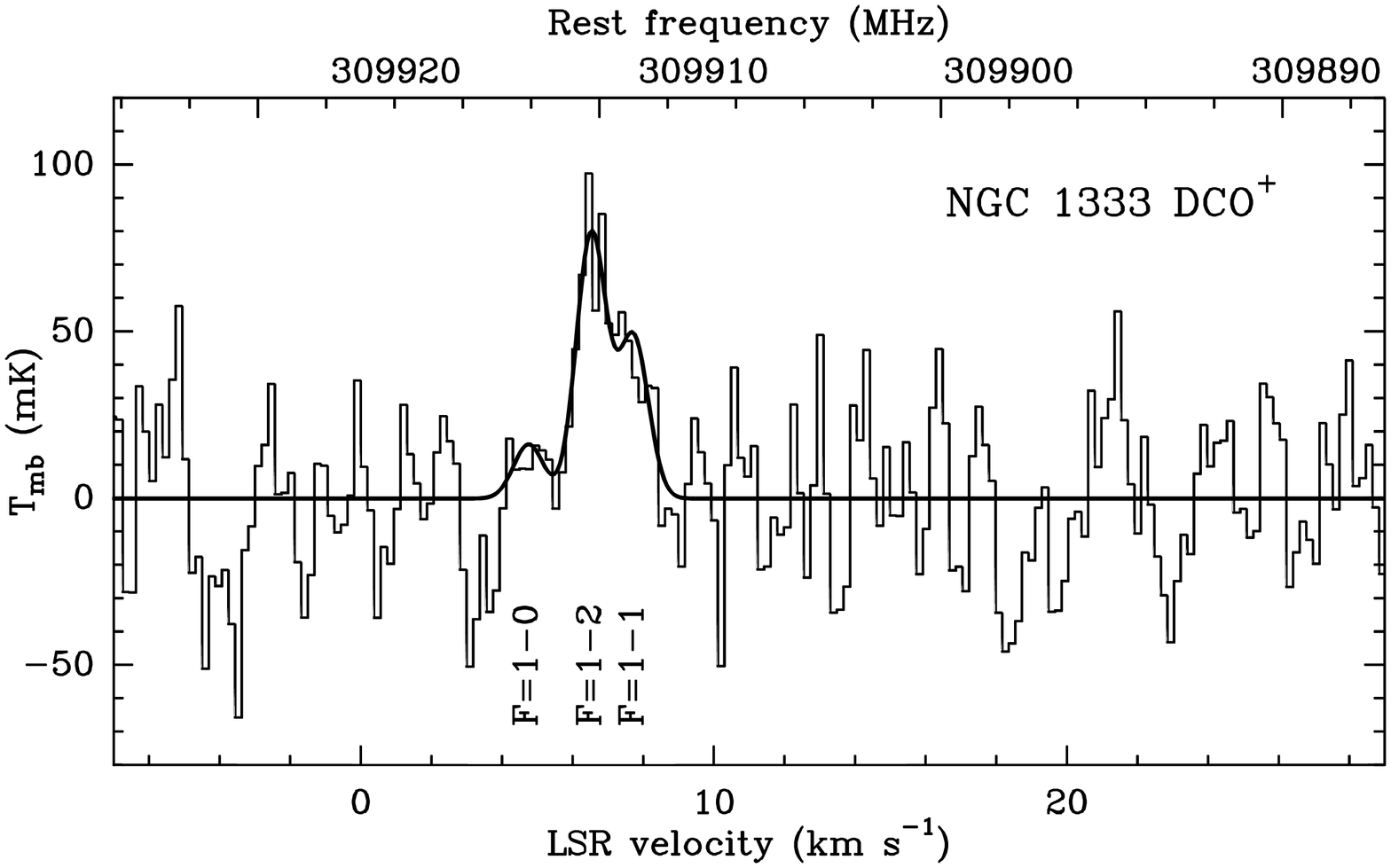}}
  \caption{Top: Spectrum of NGC 1333 IRAS 4A near 309.9 GHz, taken with the
    CSO, with the 3-component fit described in the text overplotted
    and the expected positions of the hyperfine components of the
    \dammo\ $J_K=1_0^a\to0_0^s$ line indicated. Bottom: Spectrum taken
    at the $(+23,-06)''$ offset position, where DCO$^+$ peaks.}
  \label{fig:data}
\end{figure}

The Caltech Submillimeter Observatory (CSO) is a 10.4-m single-dish
antenna located atop Mauna Kea, Hawaii. At 310~GHz, the CSO has an
FWHM beam size of $25''$, slightly larger than the diffraction limit.
Initial observations of \dammo\ $J_K=1_0^a\to0_0^s$ toward NGC 1333
IRAS 4A ($\alpha_{2000}=03^h29^m10^s.3$,
$\delta_{2000}=+31^\circ13'32''$) were carried out December
5-8, 2001. Additional data were taken January 24, 2002, with the local
oscillator frequency shifted to verify the sideband origin of detected
features.  Weather conditions were average with zenith opacities of
$\approx$0.1 at 225~GHz. An IF frequency of 1594~MHz was used to put
the $J_K=1_0^s\to0_0^a$ transition in the image sideband. To subtract
atmospheric and instrumental background, positions $240''$ offset were
observed using the chopping secondary mirror. The main beam efficiency
at the time of observations, measured through observations of Saturn,
was 64\%, using a planetary brightness temperature of $135$~K at
310~GHz.

\begin{table*}[t]
  \caption{Line parameters$^a$}
  \begin{tabular}{lccccccc} \hline
Position & $\tau$ & $V_0$ & $\Delta V$ & \tmb$\tau$ &$\int$\tmb$dV$ &
  $N$(\tex=10 K) & $N$(\tex=5 K) \\
 & & \kms\ & \kms\ & mK & mK \kms\ & 10$^{11}$~\scm\ & 10$^{11}$~\scm\ \\ \hline
IRAS 4A & 0.6(20) & 7.0(2) & 1.6(6) &  81(62) &  71(21) & 2.9(9) & 6.3(19) \\
DCO$^+$ & 0.1(27) & 6.5(1) & 1.0(2) & 149(22) & 144(43) & 5.9(18) & 12.8(38) \\ \hline
  \end{tabular}

$^a$ Numbers in brackets denote the uncertainty in units of the last decimal
  \label{tab:parms}
\end{table*}

Figure~\ref{fig:data} (top) shows the combined data from all nights,
aligned in signal frequency. This spectrum represents 151~minutes of
on-source integration with a mean \tsys\ of 521~K and a source
elevation of $44^\circ$.  A feature is detected at 309909.4~MHz, at
which frequency neither the JPL \citep[spec.jpl.nasa.gov]{pick98} nor
the CDMS \citep[www.cdms.de]{mull01} catalog lists any plausible
molecular lines other than \dammo. The $307$~GHz component is not
detected to \tmb$<14$~mK (1$\sigma$).

Data were also obtained at a position $23''$ East and $6''$ South of
the IRAS~4A position, which is where the DCO$^+$ 3$\to$2 emission
peaks (Lis et al., in preparation). The bottom panel of
Figure~\ref{fig:data} shows that the \dammo\ line is also somewhat
stronger than at the IRAS~4A position. The noise level for this
spectrum is 23~mK (1$\sigma$).

Lines of \dammo\ exhibit hyperfine structure due to coupling of the
$^{14}$N nuclear spin with the rotational angular momentum.
Splitting due to D, owing to its small quadrupole moment, is only
$\approx$200~kHz and remains unresolved in our data.  Thus, we have
fitted the observed spectrum with three Gaussian profiles with optical
depth ratios of 5:3:1 and assuming that the components have equal
widths and excitation temperatures. Such a fit has four parameters:
optical depth $\tau$, central velocity $V_0$, line width $\Delta V$
and intensity \tmb=$\tau(T_{\rm ex}-T_{\rm bg})$. We used the HFS
method inside the CLASS package. Leaving all parameters free gives the
values reported in columns 2--5 of Table~\ref{tab:parms}. For the
IRAS~4A position, fixing $V_0$ and/or $\Delta V$ to values measured in
other lines, $7.0$~and $1.2$~\kms\ \citep{blake95} gives similar
results. The fit results indicate that the line shape is consistent
with the optically thin hyperfine intensity ratio of 5:3:1, but that
the signal to noise is not high enough to constrain the optical depth.
Without taking hyperfine broadening into account, $\Delta V$ at the
IRAS~4A position would be $2.6$~\kms, much broader than other lines
from the cold component of this source \citep{blake95}. The line shape
thus confirms the assignment of the line to \dammo.

From the observed line strengths integrated between 3.5~and 10.0~\kms,
given in column~6 of Table~\ref{tab:parms}, we estimate the \dammo\
column density assuming an excitation temperature \tex=10~K.  In the
optically thin limit, but outside the Rayleigh-Jeans regime, and using
a background temperature of $T_{\rm bg}=2.7$~K, the
velocity-integrated optical depth follows from
 
$$
\int T_{\rm mb} dV = \frac{h\nu}{k}\left( \frac{1}{e^{h\nu/kT_{\rm ex}}-1}
- \frac{1}{e^{h\nu/kT_{\rm bg}}-1} \right) \int \tau dV
$$

\noindent so that

$$
\int \tau dV = \frac{c^3}{8\pi\nu^3}AN_u (e^{h\nu/kT_{\rm ex}}-1)
$$

\noindent with

$$N_u=\frac{g_uN}{Q(T_{\rm ex})}e^{-E_{\rm up}/kT_{\rm ex}}$$ 

yields the column density estimates in column~7 of
Table~\ref{tab:parms}.  Here, $Q(T_{\rm ex})$ is the partition
function $\sum_i g_i e^{-E_i/kT_{\rm ex}}$, equal to 38.3 for
\tex=10~K. The Einstein~A coefficient of $2.57\times 10^{-4}$~s$^{-1}$
follows from the dipole moment of 1.49~D \citep{lonar81}.

The assumed value of \tex\ represents a kinetic temperature at which
chemical fractionation should be efficient. However, at densities well
below the critical density of this line, $10^7-10^8$~\ccm, \tex\ will
drop below \tkin, which changes the column density estimate. As an
example, the last column of Table~\ref{tab:parms} gives the values for
\tex=5~K.

\section{Abundance of \dammo}
\label{sec:hst}

To estimate the abundance of \dammo\ we have used the Monte Carlo
radiative transfer program by
\citet[talisker.as.arizona.edu/$\sim$michiel/ratran.html]{hst00}. Lacking
auxiliary data on the DCO$^+$ position, we concentrate on NGC~1333
IRAS~4A, for which we take the temperature and density structure from
\citet{stark99}.  Between the outer and inner radii of 3100~and 10~AU,
temperatures increase from 13~to 320~K, and densities from $2\times
10^6$ to $4\times 10^{11}$~\ccm; $N$(\hh)=$3.1\times 10^{23}$~\scm\ in
a $13''$ beam, but strongly depends on beam size due to the $R^{-2}$
density distribution. The radiative transfer model for
\dammo\ includes the 30 terms up to $100$~\rcm\ above ground,
including the inversion splitting but not the hyperfine structure.
Rate coefficients for de-excitation of \ammo\ in collisions with \hh\ 
from \citet{danby88} are used, scaled to the different reduced mass of
the \dammo-\hh\ system, and augmented with the terms that are
Pauli-forbidden in \ammo, and with transitions that would be
ortho-para conversions in \ammo.  Initially, a constant abundance of
\dammo\ (relative to \hh) was assumed. The excitation of \dammo\ as a
function of radius is calculated with the Monte Carlo program. The
result is integrated over the line of sight and convolved with a
$25''$ beam. The area under the synthetic line profile matches the
observed value for \dammo/\hh$=3.2\times 10^{-12}$.

As an alternative model, the \dammo\ abundance was assumed to follow
that of \hhdp. The major chemical formation path to \dammo\ starts
with the reaction of \ammo\ with \hhdp\ and its derivatives DCO$^+$
and \nndp, and proceeds through \nhhd\ and \nddh. As a simple way to
model this behaviour, we have assumed a constant \hhdp/\dammo\ ratio.
However, this ratio would vary in the case of a varying \ammo\ 
abundance, and if the alternative route starting with the
reaction of N$^+$ with HD competes, which is slightly endothermic.  In
the absence of sufficient constraints, we keep \hhdp/\dammo\ constant.
Our two assumed \dammo\ abundance profiles could be tested indirectly
by observations of key deuterated molecules such as DCO$^+$ and \nndp.

The \hhdp\ abundance profile in NGC~1333 IRAS~4A was calculated
analytically by \citet{stark99}, using assumed values for the
cosmic-ray ionization rate ($\zeta=5\times 10^{-17}$~s$^{-1}$) and the
abundances of HD ($2.8\times 10^{-5}$) and D ($2.8\times 10^{-6}$),
and using a CO abundance of $4\times 10^{-6}$ estimated from C$^{17}$O
data. Due to the small energy difference between \hhh\ and \hhdp, the
\hhdp\ abundance is strongly peaked toward large radii where
temperatures are low. We have re-calculated the \hhdp\ abundance
profile using $\zeta=2.6\times 10^{-17}$~s$^{-1}$, the mean of the
values implied by observations of \hhh\ and H$^{13}$CO$^+$ towards
seven massive young stars \citep{zeta00}. This calculation also
includes dust radiation which the one by \citet{stark99} did not.  The
new calculations are still consistent with the measured \hhdp\ line
flux, and indicate an \hhdp\ abundance increasing from $5\times
10^{-19}$ at a radius of $10$~AU to $5\times 10^{-10}$ at $R=3100$~AU.
To model the \dammo\ data, models were run for several values of the
\hhdp:\dammo\ ratio, and agreement between observed and calculated
line flux was found for \hhdp/\dammo=46. The \dammo\ abundance at
large radii is then $1.0\times 10^{-11}$, a factor of~3 higher than
that found assuming a constant \dammo\ abundance.

\section{Chemistry of \dammo}
\label{sec:disc}

Table~\ref{tab:coldens} summarizes the measured column densities of
\ammo\ isotopomers toward NGC~1333 IRAS~4A. It is seen that
$N$(\ammo)/$N$(\nhhd) $\approx$10 and $N$(\ammo)/$N$(\dammo)
$\approx$1000; no observations of \nddh\ exist yet. The available data
suggest a trend where with each H$\to$D substitution, the column
density drops by an order of magnitude. Current models of gas-phase
chemistry, on the other hand, predict that \nddh/\dammo\ $>$
\nhhd/\nddh\ $>$ \ammo/\nhhd\ \citep{rodg01}, unless deuteron transfer
reactions are much more rapid than proton transfers.  The same trend
is expected in the case of grain surface chemistry.  However, the
measured column densities are uncertain by $\approx$30\% due to
calibration, so their ratios could be off by a factor of two and
cannot be used to rule out either mechanism.

Since a straightforward comparison of column densities may be
complicated by the differences in beam size of the data in
Table~\ref{tab:coldens}, we have determined the \ammo\ abundance
toward NGC~1333 IRAS~4A using the approach of \S~\ref{sec:hst}.  The
same temperature and density structure as in \S~\ref{sec:hst} are
used, and the original collisional rate coefficients of
\citet{danby88}.  Based on the UMIST database
\citep[www.rate99.co.uk]{umist00}, the rates of the major destruction
reactions of \ammo\ do not depend on temperature. Formation of \ammo\ 
is mainly by dissociative recombination of NH$_4^+$, the rate of which
has a $T^{-0.5}$ dependence, which in the model for NGC~1333 IRAS~4A
corresponds to a factor of~5.  This factor is dwarfed by the
exponential increase in \hhdp\, so only constant-abundance models have
been considered for \ammo. The observations of \citet{shah01} are
reproduced for \ammo/\hh$=1\times 10^{-8}$.

\begin{table}[t]
  \caption{Column densities of ammonia isotopomers towards NGC 1333
  IRAS 4A}
  \begin{tabular}{lccl} \hline
Species & $N$ & Beam & Reference \\
 & \scm\ & $''$ & \\ \hline
\ammo  & $3.1(4)\times 10^{14}$ & 74 & \citet{shah01} \\
\nhhd  & $2.2(7)\times 10^{13}$ & 90 & \citet{shah01} \\
\dammo & $2.9(9)\times 10^{11}$ & 25 & this work \\ \hline
  \end{tabular}
  \label{tab:coldens}
\end{table}

\citet{rodg01} present a chemical scheme to form deuterated ammonia in
the gas phase, which assumes that the branching ratios of dissociative
recombination are statistical and that the relevant reaction rates are
isotope-independent. Using this scheme, for \ammo/\nhhd=10
(Table~\ref{tab:coldens}), an abundance ratio of \ammo/\dammo\ of
$\approx$10000 is expected. The observed value of
$10^{-8}/10^{-11}=1000$ is inconsistent with this prediction. This
disagreement may indicate that a detailed chemical network is needed
instead of a statistical treatment. In addition, \ammo/\nhhd\ was
measured in an arcmin-sized region and may be $<10$ within the $25''$
CSO beam. Based on the observed \ammo/\dammo\ ratio, \ammo/\nhhd\
could approach unity on small scales.

In the case of surface chemistry, the observed \ammo/\dammo\ ratio
implies an atomic D/H ratio of $\approx$0.15 in the gas phase
\citep{rodg01}.  This is significantly higher than the values of
$10^{-2}-10^{-3}$ in the chemical models of \citet{rob00}.  Based on
this discrepancy and on the closer agreement of the observed
\ammo/\dammo\ ratio with the gas-phase prediction for \ammo/\nhhd=10,
we tentatively conclude that ion-molecule reactions are presently the
preferred formation mechanism of \ammo\ in NGC~1333. This mechanism
can reproduce the observed deuteration levels if deuteron transfer
reactions are much faster than proton transfers. In the future, this
conclusion should be tested through measurements of the \ammo, \nhhd,
\nddh\ and \dammo\ abundances in a larger source sample. Any
conclusion drawn from such data will be much stronger if the lines are
measured with similar beam sizes. More detailed chemical networks are
also needed.  Such a project could constrain the relative importance
of gas-phase and grain-surface deuteration as a function of
environment.

\begin{acknowledgement}
  The CSO is supported by NSF grant AST 99-80846. HSPM acknowledges
  support from the Deutsche Forschungsgemeinschaft (DFG) via grant SFB
  494.
\end{acknowledgement}

\bibliographystyle{aa}
\bibliography{Ed011}

\end{document}